\documentclass{emulateapj}
\slugcomment{Accepted and scheduled for publication  
in {\it The Astrophysical Journal}}  
\def\lax {\ifmmode{_<\atop^{\sim}}\else{${_<\atop^{\sim}}$}\fi}  
\def\gax {\ifmmode{_>\atop^{\sim}}\else{${_>\atop^{\sim}}$}\fi}  
\def\gtorder{\mathrel{\raise.3ex\hbox{$>$}\mkern-14mu
             \lower0.6ex\hbox{$\sim$}}}
\def\etal { et al. }

\begin{document}

\title{Implication of the observable  spectral cutoff energy  evolution in XTE J1550-564 }

\author{Lev Titarchuk\altaffilmark{1,2,3} and Nikolai Shaposhnikov\altaffilmark{2,4}}

\altaffiltext{1}{University of Ferrara, Physics Department, Via Saragat, 1
44100 Ferrara, Italy}

\altaffiltext{2}{Goddard Space Flight Center, NASA, 
Astrophysics Science Division, Greenbelt MD 20771; lev@milkyway.gsfc.nasa.gov}

\altaffiltext{3}{George Mason University/Center for Earth
Observing and Space Research, Fairfax, VA 22030; lev.titarchuk@nrl.navy.mil}
 
\altaffiltext{4}{CRESST/University of Maryland, Department of Astronomy, College Park MD, 20742, nikolai.v.shaposhnikov@nasa.gov}

\begin{abstract}

The physical mechanisms responsible for production of the non-thermal emission in accreting black holes (BH)
should be imprinted in the observational apperances of the power law tails in the X-ray spectra from these objects.
Different  spectral states exhibited by galactic BH binaries 
allow examination of the photon upscattering under different accretion regimes.
We revisit  the data collected by {\it Rossi X-ray Timing Explorer} ({\it RXTE}) from the BH X-ray binary  
XTE J1550-564 during two periods of X-ray activity in 1998 and 2000 focusing
%We analyze joint PCA/HEXTE spectra during both events. In particular,
on the behavior of the high energy cutoff of the power law part of the spectrum. 
For the 1998 outburst the transition from the low-hard state to the intermediate state was accompanied by a gradual decrease in the cutoff energy which then showed an abrupt reversal to a clear increasing trend as the source evolved to the very high and high-soft states. The 2000 outburst showed only the decreasing part of this pattern. Notably, the photon indexes corresponding to the cutoff increase for the 1998 event are much higher than the index values reached during the 2000 rise transition. We attribute this difference in the cutoff energy behavior to the different partial contributions of the thermal and non-thermal (bulk motion) Comptonization in photon upscattering. Namely, during the 1998 event the higher accretion rate presumably provided more cooling to the Comptonizing media and thus reducing the effectiveness of the thermal upscattering process. Under these conditions the bulk motion takes a leading role in boosting the input soft photons. Monte Carlo simulations of the Comptonization in a bulk motion region near an accreting black hole by \citet{lt10} strongly support this scenario.

\end{abstract}


\section{Introduction}

The main feature of the X-ray spectrum observed from an accreting black hole
(BH) is a strong non-thermal component phenomenologically described by a power law with  exponential cutoff
at energies above 20 keV. The origin of this emission is attributed to multiple  Compton upscattering of
the soft photons off  energetic electrons near the compact object (Compton cloud). The 
balance between the thermal and non-thermal components in a source spectrum is the primary parameter
used to define a BH spectral state. The low hard state  (LHS) is dominated by the non-thermal power law emission with
the photon index below 2.0, while in the high soft state (HSS) the spectrum 
exhibits strong thermal component 
with the temperature around 1 keV accompanied by a weak power law tail with index above 2.0.
Transitions between these two states are  classified as the intermediate state (IS). 
During rare episodes, when an accretion rate presumably reaches 
the Eddington limit,  both thermal and power law  components appear strong in the source spectrum. These episodes are identified as the very high state (VHS) \citep[see][ for different 
flavors of BH states definitions]{rm,bell05,kw08}.  Constraining the nature 
 of the spectral energy cutoff of the power law component is essential for understanding 
the processes occurring in the immediate vicinity of the accreting BHs.

Until recently, the observational picture on the extended spectral tails in
BH X-ray binaries was based largely on the OSSE results presented by \citet{grove98}.
The main conclusion of the authors was that in the LHS the power law has a clear cutoff at
$\sim$ 100 keV, while the HSS the non-thermal part of the spectrum shows no overturn at high energies. 
The presence of the non-thermal electron population was proposed as an explanation. 
The latest observational findings, however, reveal much more detailed picture of the cutoff energy behavior.
Recently \citet{motta} pointed out the specific  pattern in the high energy 
cutoff evolution during the LHS-to-HSS spectral transition in GX 339-4. 
Namely, the authors reported the monotonic decrease of the cutoff from 120 keV  in the  LHS to 60 keV in the IS
and then its  sharp increase during the transition to the HSS.  They also pointed out
 the close connection of the cutoff energy and the fast variability in the source lightcurve.
The results by \citet{motta} along with the cutoff energy behavior in XTE J1550-564
reported here present much more detailed cutoff phenomenology due to
the RXTE frequent monitoring capability.  These results show, in particular, 
that the cutoff energy exhibits a gradual evolution from the LHS to the HSS, which is difficult to reconcile
with a simplistic no-cutoff picture for the HSS set forth by  \citet{grove98}.

An extended power law distribution of photons with respect to energy is a natural consequence 
of the repetitive scattering of the ``seed'' input photons off energetic electrons.
The thermal Comptonization is a result of upscattering when electrons have a Maxwellian energy distribution.
In this case the resulting power law has a high energy turnover at the energy 
approximately twice of the electron temperature \citep{ht95}.
While the thermal Comptonization is able to explain the X-ray spectra of accreting 
BHs in the LHS, it encounters  difficulties in the case of the HSS. 
To account for the steep extended  power-law tail one needs
to invoke geometrical configuration for the Comptonization region  with the
temperature $kT_e$ of 100-150 keV and optical depth of 0.1, which is shown to be
very unstable [see \cite{tl95}, \cite{b99}]. On the other hand, \citet{lt99}, hereafter LT99,
 showed that these steep  power-law photon 
distributions (with the photon index higher than 2.0) are produced in the convergent flow 
into compact objects when the infalling plasma temperature does not exceed 10 keV. The
characteristic feature of these spectra is an abrupt cutoff  at the energy  of the order of
 $m_e c^2=511$ keV (where $m_e$ is the electron mass and the speed of light respectively) 
 and can be modeled by an exponential cutoff with the folding energy in the range of 200$-$400 keV.

In \citet[][hereafter ST09]{st09}  we presented an observational evidence that the bulk inflow motion
phenomenon is imprinted in the correlation pattern of the spectral and variability 
properties in the form of the photon index saturation effect. It was shown that
in the IS state, when the thermal and bulk motion Comptonization (BMC) are equally important,
the index depends strongly on the mass accretion rate $\dot{M}$, 
indicated either by the spectrum  normalization or by the frequency of quasi-periodic oscillations (QPOs).
However,  in the HSS spectral index tends to saturate. 
High mass accretion rate in this state  provides an effective  cooling mechanism 
for the Compton cloud. Therefore, the BMC effect  should 
take over the thermal process.  The saturation effect was found in a number of the
galactic BH binaries including XTE J1550-564. It was also shown analytically and numerically that the slope of the 
radiation spectrum resulting  from the BMC in a cold converging inflow 
does not depend on the $\dot{M}$ for high mass accretion rates [\cite{tz98},  LT99]. 

In this Paper we study, in detail, the behavior of the energy spectrum 
observed by {\it RXTE} from the galactic BH candidate XTE J1550-564 during
two outbursts occurred in 1998 and 2000. Specifically, we concentrate on
the phenomenology of the high energy cutoff of the  power law spectral component.
These two events were significantly different in strength, with the 2000 outburst 
being much weaker and shorter in duration. During the 1998 outburst   
the source started out off  the LHS, evolving through IS and reached VHS during an extremely bright flare. 
The source then returned to IS and transited into HSS later on [\cite{sob00}, ST09]. 
The high energy cutoff evolution during this episodes is completely consistent with
pattern reported by \citet{motta} for the hard-to-soft transition in GX 339-4 in 2007.
 On the other hand, during the 2000 event the source did not exhibit the VHS and the peak flux 
 was five times lower than that during the 1998 outburst. We find both similarities and differences 
 in the behavior of the high energy part of the 
spectrum during the events. We explain the observed evolution
of the cutoff  folding energy $E_{fold}$  by a varying contribution of the thermal and 
the BMC processes in the upscattering the input soft photons.
We find that the cutoff phenomenology is in the excellent agreement with the scenario
when thermal and bulk motion  Comptonization  processes dominate  the LHS and HSS correspondingly,
while in the IS both processes are equally  important in boosting the low energy photons. 

In the next section we describe the data reduction and analysis. In \S 3 we 
offer the interpretation of the power-law efold  energy ($E_{fold}$) evolution in terms of
interplay between thermal and non-thermal (bulk motion) Comptonization.
Discussions and conclusions follow in  \S 4 and    \S 5 respectively. 

\section{Observations and data reduction \label{data}}

We use archival {\it RXTE} data from the HEASARC\footnote{http://heasarc.gsfc.nasa.gov/}.
For the 1998 outburst we analyzed 64 observations starting on September, 7 1998 (MJD 51063.67) and ending on November, 2 1998 (MJD 51119.0), when the source have completed a transition to the HSS. For the 2000 data we analyzed 47 pointed observations performed from April 10, 2000 (MJD 51644.5) to June 11, 2000 (MJD 51706.0), when the source 
was in the decay LHS.  The first comprehensive analysis of 1998 and 2000 outbursts from XTE J1550-564 was presented in \citet{sob00}. We refer the reader to this paper for an exhaustive account of the data and general phenomenology.

The {\it RXTE}/PCA spectra  have been extracted and analyzed in the 3.0-50 keV  energy range using the 
most recent release of PCA response calibration (ftool {\tt pcarmf v11.7}). The relevant deadtime corrections to energy spectra 
have been applied following ``The RXTE Cook Book''  recipe.The HEXTE spectra were extracted and 
analyzed in 20-250 keV energy range. The PCA and HEXTE energy spectra were modeled jointly
using XSPEC 12.0  astrophysical spectral analysis package. To fit PCA spectra we used the sum of the generic Comptonization component \citep[BMC model, see][]{bmc}   and a GAUSSIAN with the energy $\sim6.5$ keV, which is presumably related to 
the iron emission line. The model was also modified by the interstellar absorption, using the WABS model in
 XSPEC, with a hydrogen column value fixed at the Galactic value given by nH HEASARC tool \citep{nH}
and by a high energy cut-off HIGHECUT. The upper limit of 1.2 keV was applied to the width of the GAUSSIAN. 
The high energy cutoff  component accounts for the  exponential overturn of the spectrum. 
Systematic error of 1.0\% have been applied to the analyzed spectra. 

\begin{figure}[ptbptbptb]
\includegraphics[scale=0.45,angle=0]{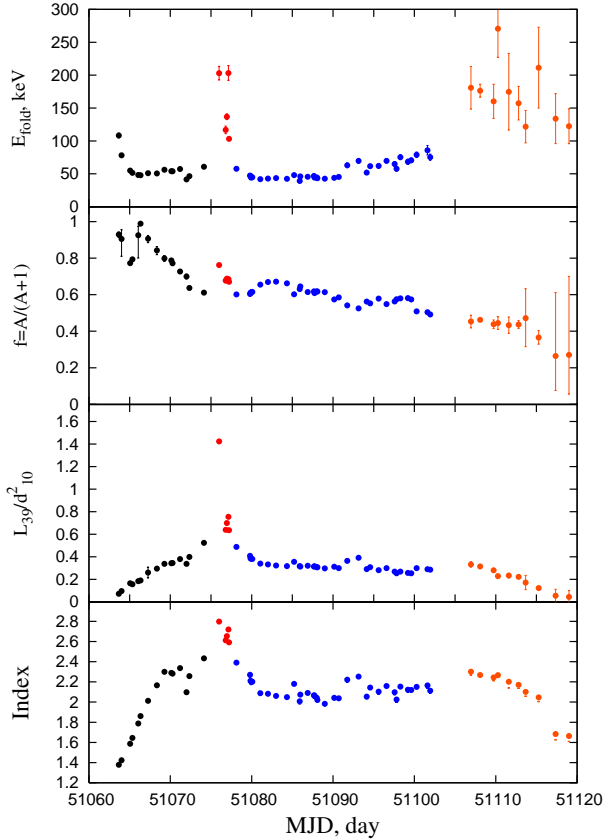}
\caption{ Evolution of spectral parameters during the rise part of the 1998 outburst of XTE J1550-564.
 LHS spectrum is shown in black, VHS is in red, IS is in blue and HSS is in orange. }
\label{evol_1998}
\end{figure}

\begin{figure}[ptbptbptb]
\includegraphics[scale=0.45,angle=0]{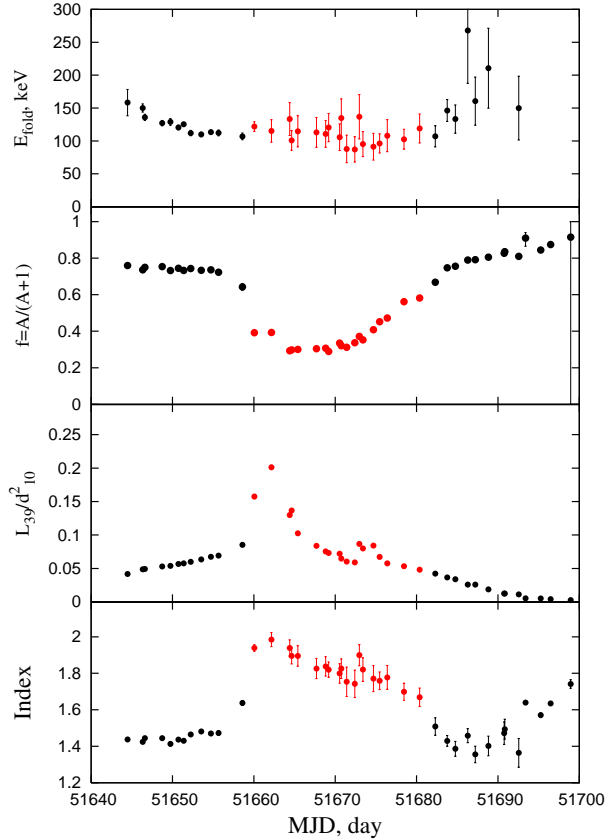}
\caption{ Evolution of spectral parameters during the 2000 outburst of XTE J1550-564. }
\label{evol_2000}
\end{figure}

The BMC model, initially intended to treat the bulk motion Comtonization, does not specify
the particular properties of the upscattering electron
gas and therefore is valid for the thermal Comptonization, as well as for the case of the hybrid thermal/non-thermal Comptonization. It describes the outgoing spectrum as a convolution of input ``seed'' black body spectrum having normalization $N_{bmc}$ and the color temperature $kT$ with a Green function for Comptonization process.
Similarly to the ordinary BBODY XSPEC model, the normalization $N_{bmc}$ is a ratio of the total input black body  luminosity 
to  the square distance
\begin{equation}
N_{bmc}=\biggl(\frac{L}{10^{39}\mathrm{erg/s}}\biggr)\biggl(\frac{10\,\mathrm{kpc}}{d}\biggr)^2.
\end{equation}  
The resulting  model spectrum is also characterized by the parameter $\log(A)$ related
to the Comptonized fraction $f=A/(1+A)$ and  the Green's  function spectral index
$\alpha=\Gamma-1$ where $\Gamma$ is the photon index. 

There are two reasons for using the BMC model. 
 First, the model by the nature of the model is applicable to the general case when
 photons gain energy not only due to thermal Comptonization but also via
dynamic or bulk motion Comptonization \citep[see LT99, ][for details]{bmc,  ST06}.
The second reason is that the BMC norm  $N_{bmc}$ is tied to  the normalization of the ``seed'' black body,
 presumably originated in the disk.  The direct correspondence of $N_{bmc}$ to the mass accretion rate in the disk  
follows from the accretion disk theory \citep[see e.g. ][]{ss73}. 
The adopted spectral model successfully describes the most spectra. The reduced $\chi^2$-statistic value
$\chi^2_{red}=\chi^2/N_{dof}$, where $N_{dof}$ is the number of degrees of freedom
for a fit, is less or around 1.0 for more than 95\% of the observations. For a small 
fraction (less than 3\%) of spectra with high counting statistic the value of
$\chi^2_{red}$ reaches 1.5. However, it never exceeds the rejection limit of 2.0.

\section{Evolution of the spectral properties during
 state transitions in XTE J1550-564\label{transitions}}

\begin{figure}[ptbptbptb]
\includegraphics[scale=0.32,angle=-90]{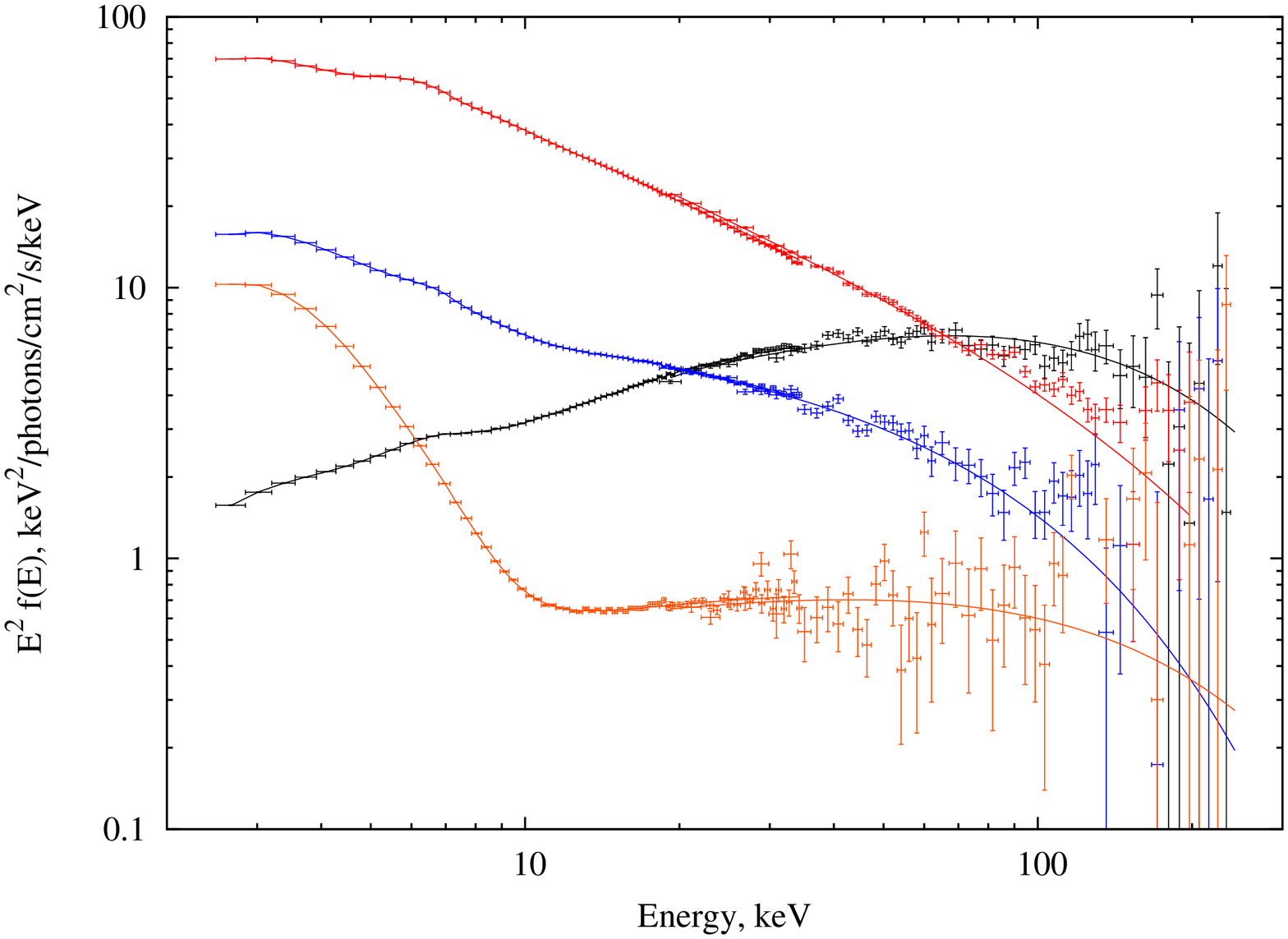}
\includegraphics[scale=0.32,angle=-90]{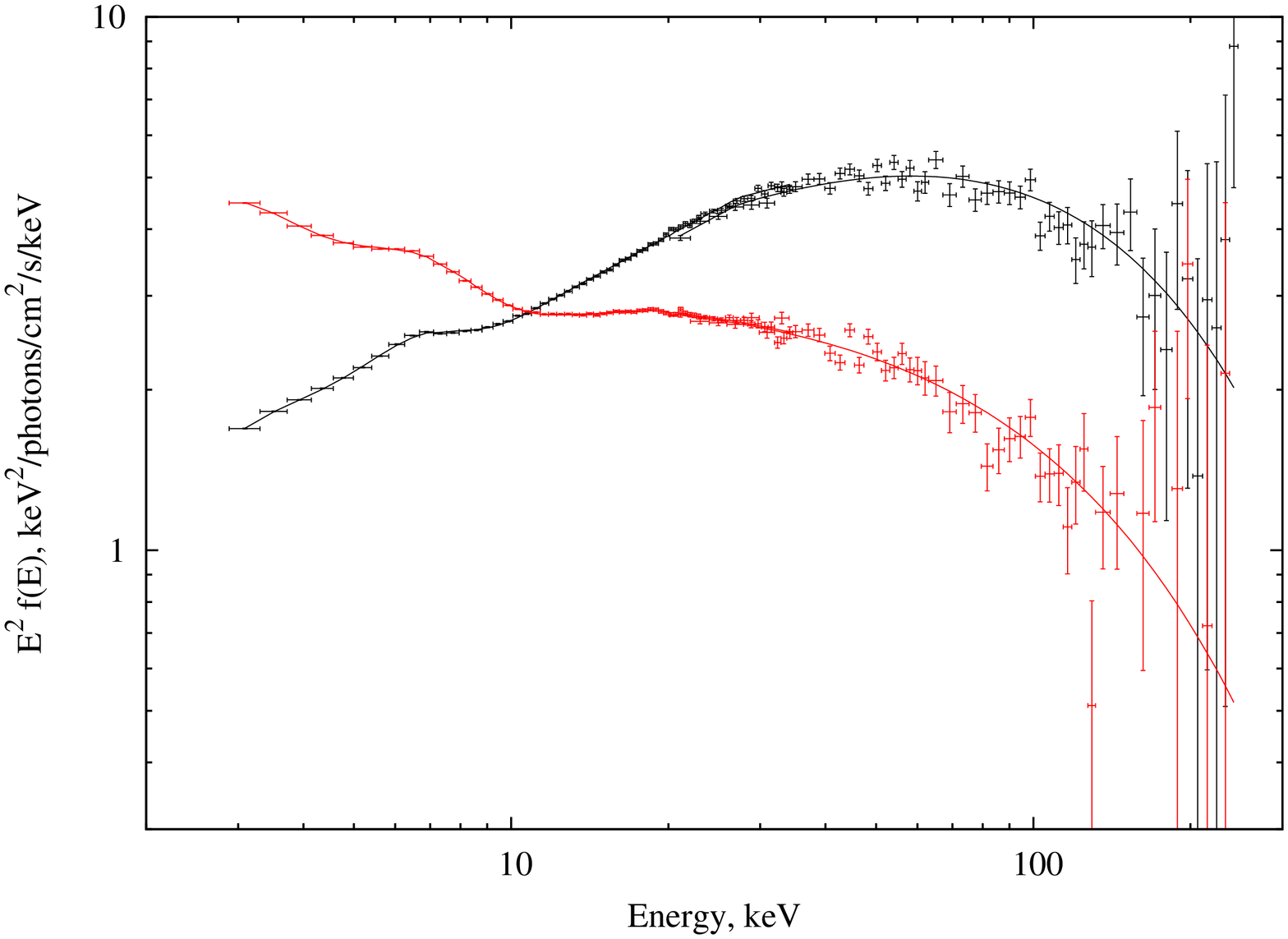}
\caption{
Four representative energy spectra during different stages of 1998 outburst from XTE J1550-564.
LHS spectrum is shown in black, VHS is in red, IS is in blue and HSS is in orange.}
\label{eeufs}
\end{figure}

Evolution of the relevant spectral parameters during the 1998 and 2000 outbursts 
is presented in  Figures \ref{evol_1998} and \ref{evol_2000} correspondingly. 
Different spectral states are separated by color.  In Figure \ref{eeufs} we illustrate the spectral evolution in XTE J1550-564  for two outbursts. The top panel presents four representative unfolded
spectra for the 1998 outburst, color coded according to the color legend of Figure \ref{evol_1998}, 
i.e. black for the LHS, blue for the IS, red for the VHS and orange for the HSS.
The 2000 event is presented in the bottom panel by two spectra for the LHS in black and the HSS in red. 

The 1998 outburst of XTE J1550-564 developed as follows. 
The outburst started on MJD 51063 and went through the initial LHS and then entered the hard IS. 
Energy $E_{fold}$  dropped from 100 keV to 50 keV during this LHS-IS phase.
In Figure  \ref{evol_1998} this stage is marked by filled black circles. 
The source exhibited a strong VHS flare on 
MJD 51076, when photon index  peaked at 2.8  marked by red points. 
On the other hand $E_{fold}$  showed sharp peak up to 250 keV during the flare.
We indicate these data by  red color. After the flare
the source returned to the IS with index $\sim 2.1-2.2$ and $E_{fold}$ dropped back to ~50 keV.
% and QPO at $\sim 3$ Hz.
For the next 20 days we observe smooth evolution towards the HSS
with the photon index  increasing to  $\sim$ 2.4 (in Fig. \ref{evol_1998}  the IS-HSS transition data is shown in blue).
The cutoff folding energy $E_{fold}$ shows  a steady upward trend during 
this period, accompanied by a slow decrease of the Comptonized fraction $f$.
%This third stage of the outburst rise episode is shown in Fig. \ref{1550_evol_2} in blue color.
On about MJD 51105 the source entered the HSS, when $E_{fold}$ jumped to 150-200 keV range
(see red points in Fig.  \ref{evol_1998}). During the IS-HSS stage the source presumably went through
the strong surge of accretion. This cold accretion flow
 provided strong photon cooling for  the innermost part of the accretion flow 
 which is manifested by an increase of the photon index.

The behavior of the source during the 2000 outburst was clearly different (see Figure \ref{evol_2000}).
First, there was no VHS flare and the maximum flux reached for this 
outburst was five times less than that during the 1998 event.
The initial LHS and the hard IS are indicated by black data points from MJD 51644 to MJD 51659.
During this state we observe a decrease of  $E_{fold}$ energy similar to
the LHS-IS stage during the 1998 outburst, however,  reaching down only to 100 keV values apart to the 50 keV 
bottom plateau during 1998 event seen in Fig.  \ref{evol_1998}.  No apparent increase of
$E_{fold}$ is obseved for the transition  to the HSS on MJD 51660. 
Much lower overall flux observed during this outburst is probably 
indicating a lower mass accretion rate in the disk with respect to the 1998 event.
This is also reflected in the behavior of the source spectrum. Namely,
the index does not grow higher than 2.0 but $E_{fold}$ stays high.
This is presumably due to   the  constant presence of the hot thermal Comptonizing media
throughout the transition. {\it The level of cold matter supply in the disk is not sufficient to 
completely  cool down the Compton Cloud in the case of 2000  outburst}.

\begin{figure}[ptbptbptb]
\includegraphics[scale=0.32,angle=-90]{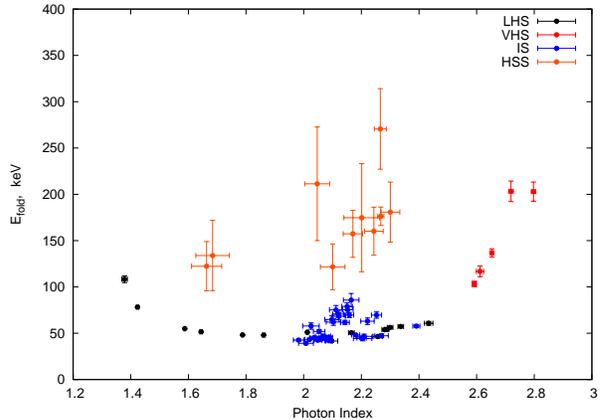}
\caption{ Efold energy $E_{fold}$ versus photon index for 1998 outburst. 
%Color legend corresponds to
%Fig \ref{evol_1998}.}
}
\label{cutoff_vs_index_1998}
\end{figure}

\section{Discussion}

LT99 studied the Comptonization of the soft radiation in the converging inflow (CI) onto 
a (BH) using Monte Carlo simulations. The fully relativistic treatment has been implemented to
reproduce the spectra. The authors show that the spectrum of the soft
state can be described as the sum of a thermal (disk) component and the convolution of some fraction of this
component with the CI  upscattering spread (Green's) function.
The latter is seen as an extended power law tail at the energies much higher than the characteristic
energy of the soft photons and plasma temperature. LT99 also demonstrate the stability of
the power-law index (the photon index   $\Gamma\sim2.8$) over a
wide range of the plasma temperatures 0$-$10 keV and mass accretion rates (higher than 2 in Eddington units) due to
upscattering and photon trapping in the CI.  However, the spectrum is
practically the same as that produced by standard thermal Comptonization when the CI plasma temperature is of the order of
50 keV (typical for the LHS) and the photon index is around 1.7. 
LT99 also demonstrate that the change of the spectral shapes from the HSS to the LHS is clearly related to
electron temperature $kT$ and optical depth of the bulk inflow $\tau$ [see  also \cite{tf04} for more details of the  $T-\tau$ dependence].   
%Moreover \cite{tf04} showed that the observable index-QPO correlation in XTE 1550-564 can be reproduced using the transition layer model. In particular, they use the fact that optical depth of the converging flow (mass accretion rate)  positively correlates with QPO frequency. 
When the mass accretion rate of the flow increases the plasma temperature decreases and thus 
the high energy cutoff $E_{fold}$ decreases until the effects of bulk motion Comptonization becomes dominant. Then $E_{fold}$  increases with mass accretion rate and weakly depends on plasma temperature $kT_e$ if it is less than 10 keV.

\begin{figure}[ptbptbptb]
\includegraphics[scale=0.32,angle=-90]{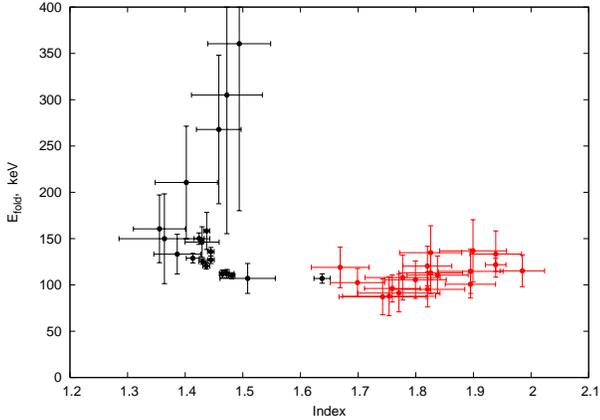}
\caption{ Efold energy $E_{fold}$ versus photon index for 2000 outburst. Color legend corresponds to
Fig \ref{cutoff_vs_index_1998}.
}
\label{cutoff_vs_index_2000}
\end{figure}

Observationally this bulk  motion effect can be seen as a non-monotonic  behavior  $E_{fold}$ versus index $\Gamma$. This is exactly what is observed during the initial LHS and the VHS flare of the 1998 outburst in Figure \ref{cutoff_vs_index_1998}). 
Namely, the black and red data points follow clear non-monotonic curve starting from $E_{fold}$~150 keV and the index of 1.4 in LHS, decreasing to the minimum $E_{fold}$ of 50 keV for the index of 2.0 at the start of the VHS flare  and reaching
the $E_{fold}$ of 200 keV and the index of 2.9 at the VHS peak. Then the source returned back to the values
reached at the minimum. During the IS which followed after the VHS flare the source took a different path marked
by smaller index values and transitioned to the HSS, shown by orange points. The $E_{fold}$ versus index
shows clearly two-parametric behavior in this case. One possible explanation for this fact is
that the VHS episode was triggered by strong surge of the matter with lower angular momentum, which then serves
as a second parameter to define the source behavior.
 
\begin{figure}[ptbptbptb]
\includegraphics[scale=0.32,angle=-90]{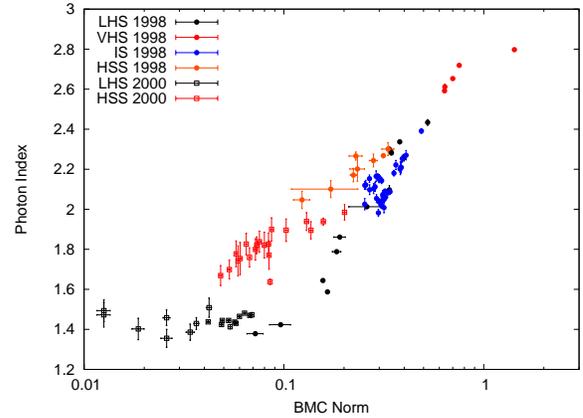}
\caption{ Observational correlation of photon index  versus BMC normalization which is proportional to disk  mass accretion. 
Filled circles used for the 1998 outburst data while the 2000 data is indicated by empty squares. 
%Color legend corresponds toFig \ref{evol_1998}.}
}
\label{index_vs_bmc_norm1998}
\end{figure}

\begin{figure}[ptbptbptb]
\includegraphics[scale=0.32,angle=-90]{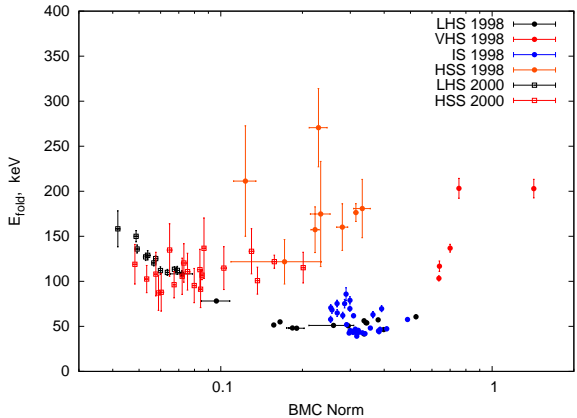}
\caption{ Efold energy $E_{fold}$   versus BMC normalization which is proportional to disk  mass accretion. We
use same point styles and color legend as in Figure \ref{index_vs_bmc_norm1998} to distinguish data for different
outbursts and BH spectral states. 
%Color legend corresponds toFig \ref{evol_1998}.}
}
\label{cutoff_vs_bmc_norm1998}
\end{figure}

During the 2000 outburst (Fig.  \ref{cutoff_vs_index_2000}) we see only the first part of this $E_{fold}-\Gamma$ pattern. Namely,  $E_{fold}$ decreases and reaches down to its minimum value of 100-120 keV while 
$\Gamma$ does not exceed 2.0 (compare Figs. \ref{cutoff_vs_index_1998}  and  \ref{cutoff_vs_index_2000}). 
It is interesting to note that the HSS data (orange points) for the 1998 outburst in Figure  \ref{cutoff_vs_index_1998} lie close
to the 2000 outburst data for the HSS in Figure  \ref{cutoff_vs_index_2000} (red points). This indicates that
during transition to the canonical HSS similar accretion regimes are at work for both outbursts, while during the VHS
flare a unique and more rare accretion regime occurred possibly due to a supercritical accretion of matter 
having significant fraction of the low angular momentum gas. The red VHS track on Figs. \ref{cutoff_vs_index_1998},
\ref{index_vs_bmc_norm1998} and  \ref{cutoff_vs_bmc_norm1998}
is therefore can be attributed to the higher indexes and fluxes due to strong cooling and higher energy release to
unusually high mass accretion rate during this state. 

We  see {\it the strong evidence of the bulk motion effect when the cutoff folding energy  $E_{fold}$ first decreases with  photon index  and then reaching minimum value in the range of 50-100 keV starts to increase to  about 200 keV and when  index  values start to saturate  at  $\sim2.8$.}
On the other hand,  \cite{ft10} recently show using a number of {\it Beppo}SAX observation of accreting neutron stars 
that   photon index of Comptonization spectra does not 
show strong evolution and, in fact, stays  almost constant around 2.0 from the hard to soft states of neutron star
 binaries.  
 
\begin{figure}
\includegraphics[scale=0.33,angle=-90]{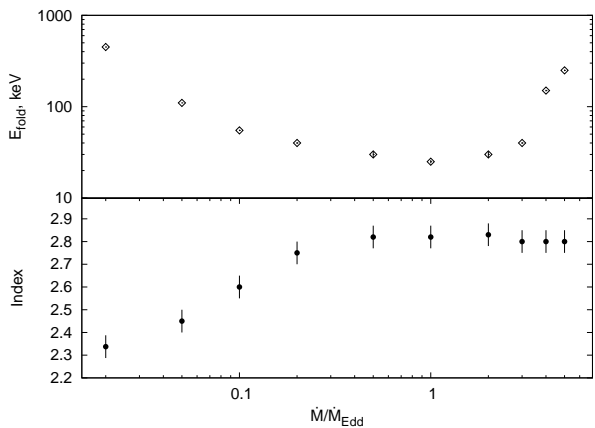}
\caption{ The cutoff folding energy $E_{fold}$ (top panel)  and the photon index  versus  mass accretion rate obtained in Monte Carlo simulations by \cite{lt10}.}
\label{cutoff_vs_mass_rate_MC}
\end{figure}

\cite{lt10} simulated spectra of  the converging flow using Monte Carlo method. If was found that
 the cutoff folding energy $E_{fold}$ of the power law component first decreases and
 then increases  as a function of  mass accretion rate $\dot m$. The cutoff energy reaches its
 minimum around $\dot{m}\sim 1$.  Also they demonstrate that  index of the emergent 
 spectrum monotonically increases for  $\dot m >0.1$ and then saturates in complete agreement with
 the observed picture shown in Figure \ref{index_vs_bmc_norm1998}, where $N_{bmc}$ is indicator of 
 the  disk mass accretion rate $\dot m$.
Because of this  monotonic behavior of index vs mass accretion rate up to the saturation level we expect 
that the same behavior pattern $\Gamma$ vs $\dot m$ and consequently $E_{fold}$ vs $\dot m$   should  be seen  in the observations. In Figure \ref{cutoff_vs_bmc_norm1998} we present 
 the observed dependence of  $E_{fold}$ as a function of $N_{bmc}$  ($\propto \dot m$).      
As one can see, the observed patterns of $E_{fold}$ and $\Gamma$ versus $N_{bmc}$ shown are strikingly similar to 
the Monte Carlo simulated folding energy and photon index evolution as a function of mass accretion rate 
[see Fig. \ref{cutoff_vs_mass_rate_MC} and  details of these simulations  in   \cite{lt10}].

\cite{grove98} reported the results of OSSE observations of seven transient black hole candidates : 
GRO J0422+32, GX 339-4, GRS 1716-249, GRS 1009-45, 4U 1543-47, GRO J1655-40, and GRS 1915-105. 
They found that the last four objects exhibit a ``power-law gamma-ray state'' with a soft spectral index 
($\Gamma\sim2.5-3$) and no evidence for a spectral break.  For GRO J1655-40, the lower limit on the break 
energy was found to be  690 keV.  Although \cite{grove98}  suggested  that the HSS  spectra detected by OSSE 
are consistent with bulk-motion Comptonization in the convergent accretion flow, \cite{zdz01} reinterpreted 
the same data and ruled out the bulk Comptonization as an origin of the HSS spectra because the emergent 
spectra extended to energies up to 700 keV. 

Thus, the question is which instrument more accurately describes the phenomenology  of the high energy cutoff, OSSE or 
{\it RXTE}/HEXTE.  Answer to this question is crucial for understanding and interpretation of the BH spectral signatures.  
In that respect we note that the typical  OSSE spectrum require exposure time of about $10^5$ s, i.e more than a day,
whereas a typical PCA/HEXTE observation lasts a few kiloseconds,  two orders of magnitude shorter than a usual OSSE exposure. OSSE observed GRO J1655-40 during the 1996 outburst VHS when the source was very variable on time scales of hours and longer. In this case the long accumulation time can result in a biases in the observed spectral shape.
We suggest that the presence of the extended power law up to $\sim 700$ keV in OSSE spectra can be a result of the long accumulation time scale when specific details of the spectra can be biased particularly at high energies.

Despite the fact that HEXTE is not sensitive above 300 keV, it is able to sample the source spectrum 
with much more detailed temporal resolution. Our analysis of the PCA/HEXTE from XTE 1550-564 indicates that VHS spectra do show exponential turnover at energies about 200 keV  (see Fig. \ref{cutoff_vs_index_1998}). Moreover, 
our results clearly show that the cutoff energy changes gradually from the LHS through the IS towards the HSS.  
%Based on their analysis  
It is worth noting that \citet{motta} concluded that the cutoff power law in the PCA/HEXTE spectrum of GX 339-4 is most likely 
due to one spectral component. These facts indicate strongly 
%that one spectral component showing a cutoff power law 
%would turn into unconstrained power law. 
%Thus we rather incline to believe
 that HEXTE more be more reliable source of information on the hard  tails of X-ray spectra, at least,  up to 300 keV  than OSSE. {\it In other words, one  can definitely see an exponential rollover at energies about 200-300 keV  in HEXTE data, 
 it may not be seen in OSSE data due to lower counting statistics and bias introduced by long exposures.}

In fact, there are more physical arguments in favor of the PCA/HEXTE versus OSSE  observations of the
hard X-ray tails in BH X-ray binaries. Specifically, the dynamical time scale which is related to the magneto-acoustic 
oscillations of the Compton cloud (CC) is $t_d\sim 2L_{cc}/V_a$ where $L_{cc}$ is the CC size and $V_a$ is the magnetoacoustic velocity [see e.g. \cite{ts05}]. 
With the assumption that the characteristic CC size $L_{cc}$   in the HSS is  of the order of 
$(5-10) (3 R_S) \sim 10^8(m/10)$,  where
$R_S=3\times 10^5m$ is the Schwarchild radius, $m$ is a BH mass in solar units, 
$V_a\sim 10^7 (kT_e/1$ keV) cm s$^{-1}$, $kT_e$ is Compton cloud temperature, we obtain 
that  $t_d\sim 20[(m/10)/(kT_e/1$keV)]  s. Thus, the dynamical time scale of the  Compton cloud  $t_d$  is only one order magnitude shorter than  the PCA/HEXTE spectrum accumulation time of $\sim 10^3$ s and we rather believe that the PCA/HEXTE  spectra including its turnover more precisely describe the shape of the high/soft emergent spectra than that by the OSSE spectra averaged over 2 magnitudes longer periods. 

\section{Conclusions} 
 \label{summary}
 
We present further observational evidence supporting
the theory of the bulk motion (converging) flow near accreting black holes.
We show that when sufficient cooling is provided by 
the mass supply from the donor star, the Comptonizing media
is completely cooled down and the origin of the extended cutoff power law
is due to non-thermal bulk motion process. The energy of the high energy
cutoff observed during 1998 outbursts from XTE J1550-564 (as well as  during 2007 outburst from GX 339-4 
reported by \citet{motta}) behaves in striking agreement with the bulk motion scenario.

Combined with the previously reported effect of index saturation  in BH X-ray binaries \citep{st09} 
the cutoff energy behavior provides robust observational signature of the
bulk motion region near the accreting object. As a direct consequence of the
specific drain properties of the BH, this signature presents the most direct observational
evidence of the existence of the astrophysical black holes.

The {\it RXTE} data for this work was aquired through HEASARC. Authors  acknowledge the support of
this research by NASA grant NNX09AF02G.

\end{document}